\documentclass[]{acmart}
\AtBeginDocument{%
  \providecommand\BibTeX{{%
    \normalfont B\kern-0.5em{\scshape i\kern-0.25em b}\kern-0.8em\TeX}}}

\copyrightyear{2022}
\acmYear{2022}
\setcopyright{rightsretained}
\acmConference[FAccT '22]{2022 ACM Conference on Fairness, Accountability, and Transparency}{June 21--24, 2022}{Seoul, Republic of Korea}
\acmBooktitle{2022 ACM Conference on Fairness, Accountability, and Transparency (FAccT '22), June 21--24, 2022, Seoul, Republic of Korea}\acmDOI{10.1145/3531146.3533099}
\acmISBN{978-1-4503-9352-2/22/06}

\newcommand\word[1]{\textit{#1}}

\usepackage{array}
\newcolumntype{C}[1]{>{\centering\arraybackslash}p{#1}}

\begin{document}

\title{Measuring Representational Harms in Image Captioning}



\author{Angelina Wang}
\affiliation{
  \institution{Princeton University}
  \city{Princeton, NJ}
  \country{USA}
  }
\email{angelina.wang@princeton.edu}

\author{Solon Barocas}
\affiliation{%
  \institution{Microsoft}
   \city{New York, NY}
   \country{USA}
}
\email{solon.barocas@microsoft.com}

\author{Kristen Laird}
\affiliation{%
  \institution{Microsoft}
  \city{New York, NY}
  \country{USA}
}
\email{kristen.laird@microsoft.com}

\author{Hanna Wallach}
\affiliation{%
  \institution{Microsoft}
\city{New York, NY}
\country{USA}
}
\email{wallach@microsoft.com}


\begin{abstract}
  Previous work has largely considered the fairness of image
  captioning systems through the underspecified lens of ``bias.'' In
  contrast, we present a set of techniques for measuring five types of
  representational harms, as well as the resulting measurements
  obtained for two of the most popular image captioning datasets using
  a state-of-the-art image captioning system. Our goal was not to
  audit this image captioning system, but rather to develop
  normatively grounded measurement techniques, in turn providing an
  opportunity to reflect on the many challenges involved. We propose
  multiple measurement techniques for each type of harm. We argue that
  by doing so, we are better able to capture the multi-faceted nature
  of each type of harm, in turn improving the (collective) validity of
  the resulting measurements. Throughout, we discuss the assumptions
  underlying our measurement approach and point out when they do not
  hold.\looseness=-1
\end{abstract}




\maketitle
\section{Introduction}
Image captioning refers to the task of generating a single sentence to
describe the most salient aspects of an
image~\cite{vinyals2015showandtell, you2016semantic, lu2017look,
anderson2018caption}. It is an especially challenging task that
combines computer vision and natural language processing. With
advances in both areas due to the advent of deep learning, image
captioning systems have improved significantly, leading to a variety
of real-world applications, such as generating image descriptions for
blind and low-vision users.\looseness=-1

At the same time, there are growing concerns about the fairness of
image captioning systems and the various harms they can
cause. These concerns have been considered in previous research
through the underspecified lens of
``bias''~\cite{hendricks2018snowboard, tang2020mitigating,
bhargava2019caption, zhao2021caption}. In contrast, we present a set
of techniques for measuring representational harms---that is, harms
that occur when some social groups are cast in a less favorable light
than others, affecting the understandings, beliefs, and attitudes that
people hold about these social
groups \cite{barocas2017problem}---caused by image captioning
systems. To do this, we use a taxonomy of five types of
representational harms introduced by
\citet{katzman2021tagging} in the context of image tagging.

We propose multiple measurement techniques for each type of harm. We
argue that by doing so, we are better able to capture the multi-faceted
nature of each type of harm, in turn improving the (collective)
validity of the resulting measurements. Our measurement techniques
vary in their intended uses. Some are best viewed as mechanisms for
surfacing when harms might exist (i.e., as an entry point for further
exploration) by providing overinclusive, upper bounds, while others
are more narrowly targeted and yield measurements that can be taken at
face value. However, in all cases, they are intended to be faithful to
the underlying types of harms. Because any
measurement approach necessarily involves making assumptions that may
not always hold, we aim to be as transparent as possible about our
assumptions throughout. We present measurements obtained using our
measurement techniques for two image captioning datasets using a
state-of-the-art image captioning system. Our goal
was not to audit this image captioning system, but rather to develop
appropriate measurement techniques, in turn providing us with an
opportunity to reflect on the many challenges involved.\looseness=-1

Despite our best efforts to develop normatively grounded measurement
techniques that are well-tailored to the unique characteristics of
image captioning, our analysis demonstrates that this is a very
difficult task and that numbers never tell the full story. There are
many ways to measure representational harms and although we chose to
use the specific techniques described in this paper, there are many
other techniques we could have used instead. As a result, our choices
should not be viewed as definitive, but rather an illustration of what
it looks like to attempt to measure representational harms caused by
image captioning systems. We therefore hope that our work serves as an
entry point for others to build on when developing measurement
techniques, especially in the context of image captioning.

In the next section, we give a brief overview of image captioning,
explaining how it works and what makes it unique, as well as
summarizing previous research on the fairness of image captioning
systems. In Section~\ref{sec:mechanism}, we describe our approach to
measuring representational harms caused by image captioning
systems. After that, in Section~\ref{sec:empirical}, we present our
measurement techniques, as well as the resulting measurements obtained
for two image captioning datasets using a state-of-the-art image
captioning system. Finally, we provide a short discussion in
Section~\ref{sec:discussion} before concluding in
Section~\ref{sec:conclusion}.\looseness=-1

\section{Image Captioning}
\label{sec:background}
\subsection{How image captioning works}

\subsubsection{Task}

\begin{figure*}[t!]
  \centering
  \includegraphics[width=\linewidth]{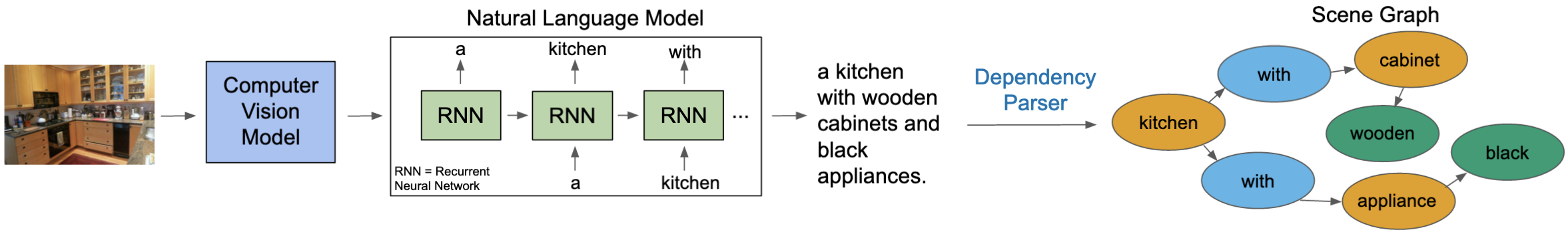}
  \caption{The operation of an image captioning system for an image of
    a kitchen. The image is fed through a computer vision model, which
    outputs visual features and, in some cases, labels that capture
    salient aspects of the image. These visual features and labels are
    then fed through a natural language model, which autoregressively
    generates a caption for the image. Finally, the caption may be fed
    through a dependency parser to generate a scene graph. (We use
    scene graphs in our approach to measuring representational
    harms.)}
  \label{fig:caption_model}
  \Description{A sequential diagram that first shows an image of a
    kitchen scene, which feeds into a square that says "computer
    vision model", and then a block that says "natural language model"
    and is composed of Recurrent Neural Network components. This then
    outputs "a kitchen with wooden cabinets and black appliances. This
    is then fed through a dependency parser that generates a scene
    graph, which has the object of kitchen pointing to two verbs of
    "with." The first "with" points to the object "cabinet" which has
    the adjective "wooden," and the second "with" points to an object
    "appliance" with the adjective "black."}
\end{figure*}

Image captioning refers to the task of generating a single sentence to
describe the most salient aspects of an
image~\cite{vinyals2015showandtell, you2016semantic, lu2017look,
  anderson2018caption}. In turn, this involves identifying what is
depicted in the image and generating coherent, descriptive text. For
example, Figure~\ref{fig:caption_model} depicts the operation of an
image captioning system for an image of a kitchen. The resulting
caption only mentions that the kitchen has wooden cabinets and black
appliances, omitting all other information.\looseness=-1

\subsubsection{Datasets}
\label{sec:datasets}

\begin{figure*}[t!]
  \centering
  \includegraphics[width=.98\linewidth]{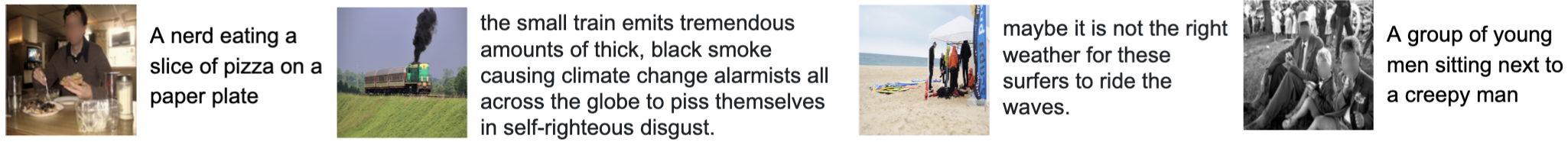}
  \caption{Examples of low-quality human-generated captions from COCO.}
  \label{fig:ex_human_captions}
  \Description{Four images with the following captions: "A nerd eating
    a slice of pizza on a paper plate", "the small train emits
    tremendous amounts of thick, black smoke causing climate change
    alarmists all across the globe to piss themselves in
    self-righteous disgust.", "maybe it is not the right weather for
    these surfers to ride the waves.", and "A group of young men
    sitting next to a creepy man."}
\end{figure*}

Common Objects in Context (COCO)~\cite{chen2015cococap} and Conceptual
Captions (CC)~\cite{sharma2018cc} are two of the most popular image
captioning datasets. Although these datasets are both intended to
support the task described above, they were created using very
different processes. COCO~\cite{lin2014coco} consists of 123,287
images from Flickr. Each image is paired with five captions and a rich
set of annotations consisting of the bounding boxes for 80 object
types. The captions were obtained from humans using Amazon Mechanical
Turk and instructions like ``describe all the important parts of the
scene,'' ``do not describe things that might have happened in the
future or past,'' and ``do not give people proper
names''~\cite{chen2015cococap}. Despite this human-driven annotation
process, the resulting captions have many quality issues, as
illustrated in Figure~\ref{fig:ex_human_captions}.\footnote{To preserve privacy, we have blurred all faces depicted in images.} Meanwhile,
CC~\cite{sharma2018cc} consists of 3.3 million images scraped from the
web. Each image is paired with a single caption that was obtained from
the image's alt-text HTML attribute, rather than from humans.\footnote{Alt-text is short for
  ``alternative text'' and refers to descriptive text, usually written
  by a human, that is intended to convey what is depicted in an
  image.}  Specifically, each image's alt-text was extracted and fed
through a data cleaning pipeline that, among other things, discarded
images with pornographic or profane alt-text and used Google Knowledge
Graph Speech and Named Entity Recognition to replace entities (e.g.,
actors' names) with their entity labels (e.g., \word{actor}). The
resulting captions are highly variable in their quality (e.g., ``video
3840x2160 -classic colored soccer ball rolling on the grass field and
stops.'',
``make a recipe that will please the whole family with this recipe!'')
because there are no enforced quality standards for alt-text.\looseness=-1

\subsubsection{Models}
\label{sec:models}

As depicted in Figure~\ref{fig:caption_model}, an image captioning
system consists of two models: a computer vision model and a natural
language model. In the case of the VinVL image captioning
system~\cite{zhang2021vinvl}, which we focus on in our analysis, an
image is first fed through the computer vision model, which outputs
visual features and labels that capture salient aspects of the image;
these visual features and labels encode the same information using
different representations. After this, the visual features and labels
are then fed through the natural language model, which
autoregressively (i.e., by conditioning the generation of each
successive word on all previously generated words) generates a caption
for the image. Prior to feeding the labels through the natural
language model, they are converted to word
embeddings. State-of-the-art image captioning systems like VinVL
typically use neural networks and transformer
architectures~\cite{you2016semantic, lu2017look, anderson2018caption,
  li2020oscar}.\looseness=-1

\subsubsection{Training}
\label{sec:training}

The computer vision model is pretrained to extract meaningful visual
features using one or more image datasets, while the natural language
model is pretrained to extract meaningful language features using one
or more text datasets. These datasets can include the dataset that
will eventually be used to train the image captioning
system.\looseness=-1

Most image captioning systems are trained using gradient descent with
a maximum likelihood objective, although this approach has been shown
to generate less diverse captions than GAN-based losses or
humans~\cite{dai2017natural, vanmiltenburg2018diversity}. Less diverse
captions mean that many different images may end up with the same
generic caption (e.g., ``A person playing tennis.'') making it
impossible to discriminate between these images from their captions
alone. As a result, the image captioning community is moving toward
other training approaches that are able to generate more diverse
captions~\cite{luo2018discriminability, shetty2017adversarial,
  dai2017natural, shi2021enhancing}.

\subsubsection{Evaluation}
\label{sec:evaluation}

Evaluating image captioning systems has proven to be extremely
challenging, meaning that there are few metrics that align well to
human judgments~\cite{cui2018evaluate}. As a result, most systems are
evaluated using a variety of different metrics---typically
BLEU~\cite{papineni2002bleu}, METEOR~\cite{banerjee2005meteor},
ROUGE~\cite{lin2004rouge}, CIDEr~\cite{vedantam2015cider}, and
SPICE~\cite{anderson2016spice}. The first four of these metrics
evaluate captions by considering the n-grams that compose them,
thereby capturing properties like fluency. SPICE instead captures
semantic quality by comparing scene graphs. Specifically, SPICE uses a
dependency parser (as shown in Figure~\ref{fig:caption_model}) to
extract three types of tuples from each caption: (object), (object,
attribute), and (subject, relationship, object). These tuples can then
be assembled into a scene graph by turning each component of each
tuple into a node.\looseness=-1

\subsubsection{Applications}

Applications of image captioning systems include indexing search
results~\cite{iyer2018searchengine}, describing images using virtual
assistants~\cite{aneja2018convolutional}, and helping non-experts
interpret domain-specific images (e.g., a medical
X-ray)~\cite{ayesha2021medicalimage}. However, by far the most
commonly mentioned application is generating image descriptions for
blind and low-vision users.\footnote{Although it falls outside the
  scope of this paper, we note that there is a worrisome disconnect
  between technical research on image captioning, which often uses
  accessibility as a motivation, and usability research focused on the
  real-world value of image captioning~\cite{stangl2020vision,
    salisbury2017scalable, mack2021alttext, macleod2017caption,
    stangl2021onesize}. For example, \citet{wu2017fb} noted that the
  captions found in datasets like COCO and CC often do not meet the
  stated needs of blind and low-vision users. Indeed, we found that
  neither dataset's captions contain proper names and both datasets'
  captions contain phrases like ``picture of'' and ``image of''---all
  of which violate quality standards for
  alt-text~\cite{heinrich2020alt}. From a fairness perspective, the
  use of accessibility as a motivation to justify investments in image
  captioning is especially troubling if those investments do not
  result in improvements to the real-world value of image captioning
  for blind and low-vision users.}

\subsection{What makes image captioning unique}

We now describe the characteristics of image captioning that separate
it from other machine learning tasks. Although image captioning is
similar to object detection and image tagging---both tasks that have
been the subject of previous research on
fairness~\cite{buolamwini2018gendershades, devries2019everyone,
  barlas2020see}---it also differs from them in several important
ways. First, object detection and image tagging aim to identify all
entities present in an image. In contrast, image captioning focuses on
only the most salient aspects of an image.\footnote{Although it is
  possible to conceive of image tagging in a way that involves tagging
  an image with only the most salient tags, in practice, this is often
  implemented as tagging an image with only the most confident
  tags. For example,
  \url{https://issuetracker.google.com/issues/117855698?pli=1} shows
  that even though Google's Vision API claims to report separate
  ``topicality'' (i.e., relevancy) and ``score'' (i.e., confidence)
  values, the same value is reported for both.\looseness=-1} Second,
although object detection and image tagging systems are not restricted
to identifying objects, they usually focus on objects rather than
attributes and relationships. Moreover, when they do generate
adjectives and verbs, these are rarely associated with particular
entities. In contrast, image captioning systems must generate all
parts of speech, including adjectives, verbs, and prepositions;
adjectives and verbs must therefore be associated with particular
entities. Finally, object detection and image tagging systems use
predefined sets of labels or tags. In contrast, image captioning
systems typically use open-ended vocabularies meaning that they can
generate any word.\looseness=-1

These characteristics make image captioning susceptible to a unique
set of fairness-related harms. First, by focusing on only the most
salient aspects of an image---an inherently subjective choice---there
is considerable room for differential treatment of different social
groups. Second, by generating all parts of speech and associating
adjectives and verbs with particular entities, some adjectives and
verbs may be systematically associated with some social groups but not
others. Third, using open-ended vocabularies makes it especially
challenging to anticipate all of the harms that may be caused by image
captioning systems. Finally, we note that the multimodal nature of
image captioning means that fairness-related harms can be caused by a
system's computer vision model, natural language model, or both
operating together (i.e., the system as a whole). For example, a
computer vision model may only treat a soccer ball as salient if it is
pictured with a masculine-presenting person; a natural language model
that starts a caption with \word{A woman} may reproduce gender
stereotypes; and a system may only mention a paintbrush if it is held
by a person with a light skin tone.\looseness=-1

\subsection{Previous research on fairness}

Previous research has largely considered the fairness of image
captioning systems through the underspecified lens of ``bias''---a
problem discussed by \citet{blodgett2020nlpbias} in the context of
natural language processing. In addition, many previously proposed
measurement techniques are not specific to image captioning and its
unique set of fairness-related harms. For example, many papers have
narrowly focused on whether image captioning systems can accurately
predict the (binary) genders of people depicted in
images~\cite{hendricks2018snowboard, tang2020mitigating}. \citet{bhargava2019caption} additionally considered
whether these predicted genders influence other aspects of caption
generation. \citet{zhao2021caption} branched out from gender
prediction to investigate differences in caption generation for images
of people with different skin tones. However, although they uncovered
a variety of differences, they stopped short of pinpointing the
fairness-related harms that might be caused by these differences. In
contrast, \citet{vanmiltenburg2016stereotype} focused specifically on
one type of fairness-related harm---stereotyping---and created a
taxonomy of how harms of this type might arise. However, they did not
investigate measurement techniques, limiting the taxonomy's
utility. Finally, other researchers have focused on the fairness of
image captioning datasets. For example, \citet{birhane2021multimodal}
identified a range of problematic content in the LAION-400M
dataset~\cite{schuhmann2021laion}, including not-safe-for-work images,
\citet{vanmiltenburg2018people} studied the adjectives used to
describe people depicted in the Flickr30K dataset, and
\citet{otterbacher2019people} investigated crowdworkers' tag choices
for a controlled set of images.\looseness=-1

\section{Measurement Approach}
\label{sec:mechanism}

\subsection{Stakeholders}

The stakeholders that could be harmed by an image captioning system
include the people depicted in images and the people to whom generated
captions are presented. We focus on harms that affect the people
depicted in images.\looseness=-1

\subsection{Types of representational harms}
\label{sec:types of harms}

We use a taxonomy of five types of representational harms introduced by
\citet{katzman2021tagging} in the context of image tagging. The first
of these types is \textit{denying people the opportunity to
  self-identify}, which occurs when identity categories are imposed on
people without their consent or, in some cases, knowledge. The second
is \textit{reifying social groups}, which occurs when relationships
between specific visual characteristics and social groups are
presented as natural, rather than historically and culturally
contingent. The third is \textit{stereotyping}, which occurs when
oversimplified beliefs about social groups reproduce harmful social
hierarchies. The fourth is \textit{erasing}, which occurs when people,
attributes, or artifacts associated with social groups are not
recognized. The final type is \textit{demeaning}, which occurs when
social groups are cast as being lower status and less deserving of
respect. Because these types of harms are theoretical constructs, they
cannot be measured directly and must be measured using techniques that
derive measurements from other observable
properties~\cite{jacobs2021measurement}.\looseness=-1

\subsection{Datasets and system}
\label{sec:datasets_and_system}

We focus on two of the most popular image captioning datasets,
COCO~\cite{chen2015cococap} and CC~\cite{sharma2018cc}, both of which
are described in Section~\ref{sec:datasets}, and the VinVL image
captioning system~\cite{zhang2021vinvl}, although we emphasize that
our goal was not to audit VinVL. We used 17,360 image--caption pairs
from COCO and 14,560 image--caption pairs from CC. Specifically, we
used the subset of the COCO 2014 validation set that overlaps with
Visual Genome~\cite{krishnavisualgenome} so that we could augment the
annotations from COCO with annotations from Visual Genome, as we
describe in Section~\ref{sec:points_of_comparison}; we used the subset
of the CC validation set for which we were able to generate scene
graphs. At the time of our analysis, VinVL held the top leaderboard
score for many tasks, including image captioning of COCO. It therefore
serves as a good vehicle for showcasing the kinds of fairness-related
harms that are caused by state-of-the-art image captioning
systems. VinVL's architecture is based on OSCAR~\cite{li2020oscar}, a
transformer-based system, but has an improved visual representation
from pretraining on a larger and richer dataset.\looseness=-1

\subsection{Stages of measurement}
\label{sec:points_of_comparison}

\begin{figure*}[t!]
  \centering
  \includegraphics[width=.8\linewidth]{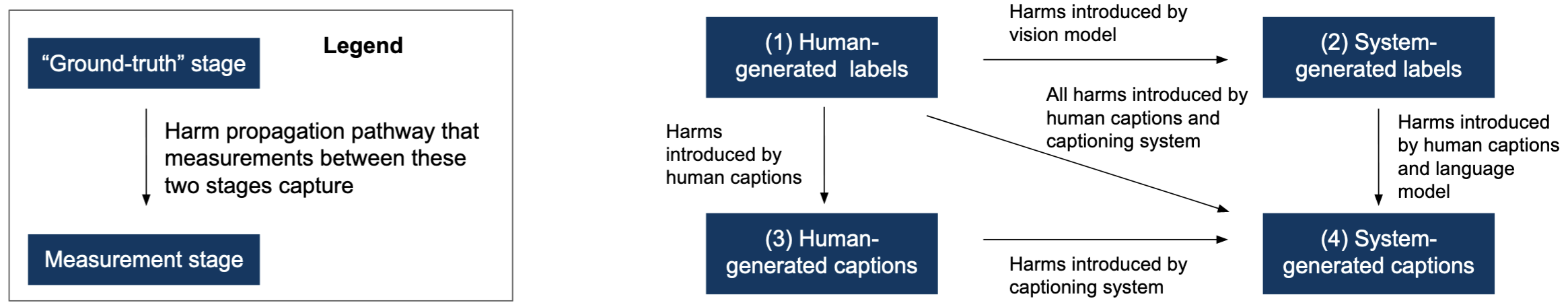}
  \caption{Our framework of four stages. For COCO, the human-generated
  labels come from a union of annotations from COCO~\cite{lin2014coco}
  and Visual Genome~\cite{krishnavisualgenome}; the system-generated
  labels come from the computer vision model that is used in the VinVL
  image captioning system~\cite{han2021image}; the human-generated
  captions come from COCO~\cite{chen2015cococap}; and the
  system-generated captions come from VinVL~\cite{zhang2021vinvl}.}
  \label{fig:scenegraph_tiers}
  \Description{Four blocks representing the four sets of scene graphs,
  with arrows going between them to indicate the harm propagation
  pathway. The four blocks are numbered as one: human annotated
  labels, two: computer vision detected labels, three: human annotated
  captions, and four: model predicted captions. The following are the
  labels on the arrows: from one to two is "harms introduced by vision
  component", from one to three is "harms introduced by human
  captions", from three to four is "harms introduced by captioning
  model", from two to four is "harms introduced by human captions and
  language component", and then from one to four is "all harms
  introduced by human captions and captioning model."}
\end{figure*}

Our measurement approach depends on a framework of four stages,
depicted in Figure~\ref{fig:scenegraph_tiers}: 1)
human-generated labels, 2) system-generated labels, 3) human-generated
captions, and 4) system-generated captions. Harms can be measured at
each of these stages in isolation or by treating one stage as ``ground
truth'' for another. For example, the presence of a demeaning word in
a caption can be measured at stage 4, without reference to the other
stages; however, a failure to describe a person depicted in an image
requires some notion of ``ground truth'' (i.e., whether there is a
person depicted in the image) and must be measured by comparing, for
example, stage 4 to stage 3. By treating different stages as ``ground
truth'' for one another, we can also better understand where harms
arise. For example, if we find evidence of a harm at stage 4
(system-generated captions), treating stage 1 (human-generated labels)
as ``ground truth,'' this harm must have arisen as a result of either
the human-generated captions or the image captioning system as a
whole; however, if we find evidence of a harm at stage 4, treating
stage 2 (system-generated labels) as ``ground truth,'' then this
harm cannot be caused by the computer vision model and must have arisen as a result of either the human-generated
captions or the natural language model; finally, if we find evidence
of a harm at stage 4, treating stage 3 (human-generated captions) as
``ground truth,'' then this harm cannot be caused by the
human-generated captions and must have arisen as a result of the
system as a whole.\looseness=-1

\textbf{Stage 1: human-generated labels:} Human-generated labels
capture everything depicted in an image, as determined by humans. To
obtain human-generated labels for COCO, we augmented its annotations
(i.e., the bounding boxes for 80 object types) with annotations from
Visual Genome~\cite{krishnavisualgenome} that include attributes and
relationships.  We also used demographic annotations collected by
\citet{zhao2021caption} that label the largest person depicted in each
image with their perceived binary gender (male or female) and skin
tone (darker or lighter). We do not have access to human-generated
labels for CC.\looseness=-1

\textbf{Stage 2: system-generated labels:} We used the labels output
by VinVL's computer vision model.\footnote{If we were interested in an
  image captioning system that did not output labels in this
  intermediary step, then provided that the computer vision model had
  been pretrained in a supervised fashion (i.e., to output labels), we
  could have instead used labels output by the pretrained computer
  vision model.}\looseness=-1

\textbf{Stage 3: human-generated captions:} We obtained
human-generated captions directly from COCO and CC.

\textbf{Stage 4: system-generated captions:} We obtained
system-generated captions for each dataset using
VinVL~\cite{zhang2021vinvl}.

The four stages described above represent information in two different
forms---labels and captions---that are difficult to compare
directly. To reconcile these differences, we therefore converted the
captions from stages 3 and 4 to scene graphs, following the approach
used by SPICE~\cite{anderson2016spice}, as described in
Section~\ref{sec:evaluation}. An example scene graph is shown in
Figure~\ref{fig:caption_model}. For COCO, in which each image is
paired with five human-generated captions, we took the union of the
scene graphs for the five captions. By using scene graphs in our
approach to measuring representational harms, we are able to focus on
the semantics of captions. Although scene graphs do not capture
meta-linguistic properties like fluency or dialect, which we
acknowledge as a limitation of our approach, we argue that syntax is
less relevant than semantics when measuring representational harms. In
addition, fluency and word choice have already been investigated
previously~\cite{zhao2021caption}.\looseness=-1

As well as converting the captions to scene graphs, we also converted
the words in the scene graphs to WordNet synsets, again to facilitate
comparisons between labels and captions. A synset is a ``grouping of
synonymous words and phrases that express the same
concept''~\cite{princeton2010wordnet}. A single word can belong to
multiple synsets. For example, \word{big} and \word{large} belong to a
synset that represents the descriptive adjective size; however,
\word{big} also belongs to 17 other synsets, including ones that
represent significance and being conspicuous in importance.
Converting words to synsets is therefore a non-trivial task. The
approach we took was to convert each word to its most common synset
that also had the appropriate part of speech (i.e., objects are
converted to nouns, attributes to adjectives, and relationships to
verbs or prepositions). If there was no synset with the appropriate
part of speech, we relaxed this constraint. Although this approach
works for the majority of words, there are cases where it can lead to
incorrect measurements of harms. For example, \word{controller} often
refers to a video game device, but its most common synset represents
an accountant. Despite these cases, converting words to synsets is
beneficial as a way to map different words to a single concept, albeit
at the cost of losing nuance; in addition, Visual Genome's annotations
are already represented as synsets. Finally, WordNet contains
hierarchies of descriptiveness via relationships between hyponyms
(e.g., \word{fork} is a hyponym of \word{utensil}) and hypernyms
(e.g., \word{color} is a hypernym of \word{red}). As we explain in
Section~\ref{sec:empirical}, we used these hierarchies when measuring
some types of representational harms.\looseness=-1

\subsection{Assumptions}
\label{sec:assumptions}

Our measurement approach necessarily involves making assumptions that
may not always hold, thereby threatening the validity and reliability
of the resulting measurements. In this section, we discuss some of our
more general assumptions; we discuss technique-specific assumptions in
Section~\ref{sec:empirical}. Because every assumption will sometimes
fail to hold, we therefore point out in Section~\ref{sec:empirical}
when our measurements are likely influenced by assumptions that do not
hold.\looseness=-1

\subsubsection{Valid and reliable external resources and tools.}

Our measurement approach involves several external resources and
tools, including word lists, WordNet~\cite{princeton2010wordnet},
NLTK's part-of-speech tagger~\cite{bird2009nltk}, SPICE's dependency
parser~\cite{anderson2016spice}, and spaCy's named entity
extractor~\cite{spacy}. As a result, one major assumption underlying
our approach is that these resources and tools are themselves valid
and reliable. For example, for some of our measurement techniques, we
needed to determine whether captions mention specific social
groups. This is particularly challenging because image captioning
systems typically use open-ended vocabularies, so it is impossible to
manually examine all possible words that could be generated to
determine which ones do indeed refer to specific social groups. To
address this challenge, we relied on word lists; in doing so, we
assumed that these word lists are exclusive (i.e., do not contain
irrelevant words), exhaustive (i.e., do not omit relevant words), and
up-to-date. In practice, though, language---especially language about
people's identities~\cite{cornell2006identities}---is continuously
evolving, so these assumptions may not hold. As another example, we
assumed that WordNet's assignments of words to synsets and
hyponym--hypernym relationships are correct and up-to-date, although
we know this may not always be the case. Indeed, previous research has
demonstrated that WordNet reflects a stagnant snapshot of
language~\cite{yang2020filter} and does not include many words that
are used to describe people's identities, such as \word{non-binary} and
\word{genderqueer}.\looseness=-1

\subsubsection{High-quality human-generated labels and captions}

Because our measurement approach treats the human-generated labels and
human-generated captions as ``ground truth,'' another major assumption
underlying our measurement approach is that the human-generated labels
and human-generated captions are high
quality\footnote{Figure~\ref{fig:ex_human_captions} contains four
  examples of of low-quality captions from COCO.} and worthy of being
treated as ``ground truth.'' This assumption is especially unlikely to
hold for the demographic annotations collected by
\citet{zhao2021caption}, which may not reflect the ways people would
like identify themselves. Moreover, by using these demographic
annotations, it is possible we have erased and mislabeled some
people~\cite{scheuerman2021nonbinary, scheuerman2020facial}---a
fairness-related harm in its own right. We also note that by treating
the human-generated captions as ``ground truth'' we are implicitly
assuming there is a ``correct'' way to caption an image (e.g., whether
a person is worth mentioning or not). In practice, because this is not
the case, some of our measurements necessarily reflect the
subjectivities inherent to the human-generated captions.\looseness=-1

\section{Measurement Techniques and Measurements}
\label{sec:empirical}
In this section, we describe the specific techniques we used to
measure the representational harms described in section~\ref{sec:types
of harms}, as well as the resulting measurements obtained for
COCO~\cite{chen2015cococap} and CC~\cite{sharma2018cc} using the VinVL
image captioning system~\cite{zhang2021vinvl}. We propose multiple
measurement techniques for each type of harm. We argue that by doing
so, we are better able to capture the multi-faceted nature of each
type of harm, in turn improving the (collective) validity of the
resulting measurements. We emphasize that although we chose to use
the specific techniques described below and in the supplementary
material,\footnote{The supplementary material is online
at \url{https://angelina-wang.github.io/files/captioning_harms_supp.pdf}.}
there are many other techniques we could have used
instead.\footnote{For example, we could have chosen to use measurement
techniques that focus on differences between social groups in the use
of abstract language (e.g., ``they are emotional'') versus the use of
concrete language (e.g., ``they have tears in their eyes''), which may
be explained by stereotyping~\cite{semin1988cognition,
beukeboom2014linguistic}.} As a result, our choices should not be
viewed as definitive. We also note that some of our measurement
techniques are best viewed as mechanisms for surfacing when harms
might exist, while others are more narrowly targeted and yield
measurements that can be taken at face value. However, in all cases,
they are intended to be faithful to the underlying types of harms. Due
to space constraints, we discuss two of the five types of
representational harms---stereotyping and demeaning---in
Sections~\ref{sec:stereotype} and~\ref{sec:demean}, respectively, and
relegate the remaining three types to the supplementary material. We
focus on stereotyping and demeaning because some of the techniques we
use to measure them are unique to image captioning. In contrast, the
techniques we use to measure the other types are more similar to techniques
proposed previously in the context of object detection or image
tagging.\looseness=-1

\subsection{Stereotyping}
\label{sec:stereotype}

As described in Section~\ref{sec:types of harms}, stereotyping occurs
when oversimplified beliefs about social groups reproduce harmful
social hierarchies~\cite{katzman2021tagging}. We propose four
techniques for measuring stereotyping in order to capture its
multi-faceted nature. The first technique focuses on cases where words
are incorrectly included in captions (i.e., false positives),
hypothesizing that these errors may be explained by stereotyping. The
second technique focuses on differences between social groups in the
objects that are correctly mentioned in captions (i.e., true
positives). Because image captioning systems describe only the most salient
aspects of an image, this technique captures a facet of stereotyping
that is unique to image captioning and does not occur in the context
of object detection or image tagging. The third and fourth techniques
are related: the third focuses on differences between social groups in
the distributions of the three types of tuples extracted from
captions, while the fourth compares these distributions across the
different stages described in Section~\ref{sec:points_of_comparison}
in order to better understand where stereotyping harms arise. We
describe the first two techniques, along with their resulting
measurements, below in Sections~\ref{sec:stereotyping_false_positives}
and~\ref{sec:stereotyping_true_positives}; the remaining two are
presented in the supplementary material as similar techniques have
been used previously in other contexts~\cite{barlas2019eye,
otterbacher2019people, wang2020revise, alvi2018removal,
kyriakou2019proprietary, barlas2019eye,
abbasi2019stereotyping}.\looseness=-1

\subsubsection{Captions that incorrectly include words}
\label{sec:stereotyping_false_positives}

We hypothesize that cases where words are incorrectly included in
captions (i.e., false positives) may be explained by stereotypes. For
example, if \word{gun} is incorrectly included the caption for an
image of someone who is Black, this is likely due to a racial
stereotype. Measuring the extent to which such cases are indeed
explained by stereotypes is challenging, however, because of the
amount of contextual and historical knowledge required. As a result,
this technique requires human interpretation and cannot be fully
automated. In other words, our first measurement technique is best
viewed as providing an overinclusive, upper bound. We therefore
propose a heuristic to rank cases where words are incorrectly included
in captions by how likely they are to be explained by stereotypes in
order to make their interpretation more tractable. The specific
heuristic we propose involves the extent of the correlation between a
word's most common synset (i.e., a ``grouping of synonymous words and
phrases that express the same concept''~\cite{princeton2010wordnet})
and a particular social group: $\max_{\mathrm{group}}[
P(\textrm{group, synset}) - P(\textrm{group})\,\cdot\,
P(\textrm{synset})]$. Words whose most common synsets are highly
correlated with some social groups are more likely to be associated
with stereotypes. For example, a case where \word{baby} is incorrectly
included in a caption is more likely to be explained by a stereotype
than a case where \word{apple} is incorrectly included. The heuristic
therefore filters the cases where words are incorrectly included in
captions using a tunable threshold (we use 0.005 in our
analysis), retaining only those cases involving a word whose most
common synset's correlation with a social group is above this
threshold. These cases are then ranked by the words' false positive
rates in order to prioritize systematic errors, which are more likely
to be explained by stereotypes, over one-offs. We emphasize that this
measurement technique involves a number of assumptions---most notably
that words whose most common synsets are highly correlated with some
social groups are more likely to be associated with stereotypes. If
this assumption does not hold, then the validity of the resulting
measurements will be threatened.\looseness=-1

To identify cases where words are incorrectly included in captions, we
focus on three scenarios. The first is where a caption includes
a \textit{non-imageable concept}, making the assumption that inferring
such a concept would require extra information that may come from a
stereotype. We used word lists to identify non-imageable concepts. For
objects, we used the non-imageable synsets in the people subtree of
WordNet~\cite{yang2020filter}; for attributes, we used those
adjectives in a list of people-descriptor
categories~\cite{vanmiltenburg2018people} that we determined to be
non-imageable (i.e., attractiveness, ethnicity, judgment, mood,
occupation or social group, relation, and state); for relationships,
we used any verb not included in Visual VerbNet\footnote{Visual
VerbNet includes verbs that relate to ``an action, state, or
occurrence that has a unique and unambiguous visual connotation,
making [them] detectable and classifiable; i.e., lay down is a visual
action, while relax is not''~\cite{ronchi2015visualverb}.} or
in \{\word{have}, \word{in}\}. The second scenario is where a caption
includes a \textit{concept that is too specific}, again making the
assumption that inferring such a concept would require extra
information that may come from a stereotype. To identify such cases
for COCO, we treated stage 1 (human-generated labels) as ``ground
truth,'' thereby assuming the human-generated labels are high quality
(e.g., if an object is labeled as \word{fruit} and not \word{apple},
we assume this is because the object is not identifiable as anything
more specific than a fruit). Because we do not have access to
human-generated labels for CC, we treated stage 3 (human-generated
captions) as ``ground truth,'' thereby assuming the human-generated
captions are high quality. As explained in
Section~\ref{sec:points_of_comparison}, this means the resulting
measurements will only reflect stereotyping harms caused by the system
as a whole and not stereotyping harms caused by the human-generated
captions. We used WordNet's hyponym--hypernym relationships to
determine whether a concept is too specific. When comparing attributes
or relationships, we only compared attributes or relationships that
refer to the same object (and subject, in the case of relationships),
as determined using Leacock Chordorow
similarity~\cite{leacockchodorow1998similarity}. The third scenario is
where a caption includes an imageable concept that is not depicted in
the image, which we refer to as a \textit{hallucination}, again making
the assumption that inferring such a concept would require extra
information that may come from a stereotype. Here too, we identified
such cases by treating stage 1 as ``ground truth'' for COCO and stage
3 as ``ground truth'' for CC. This means that our measurements for CC
do not include cases where a system-generated caption and its
corresponding human-generated caption both include a
hallucination.\looseness=-1

\begin{figure*}[t!]
  \centering
    \includegraphics[width=.95\linewidth]{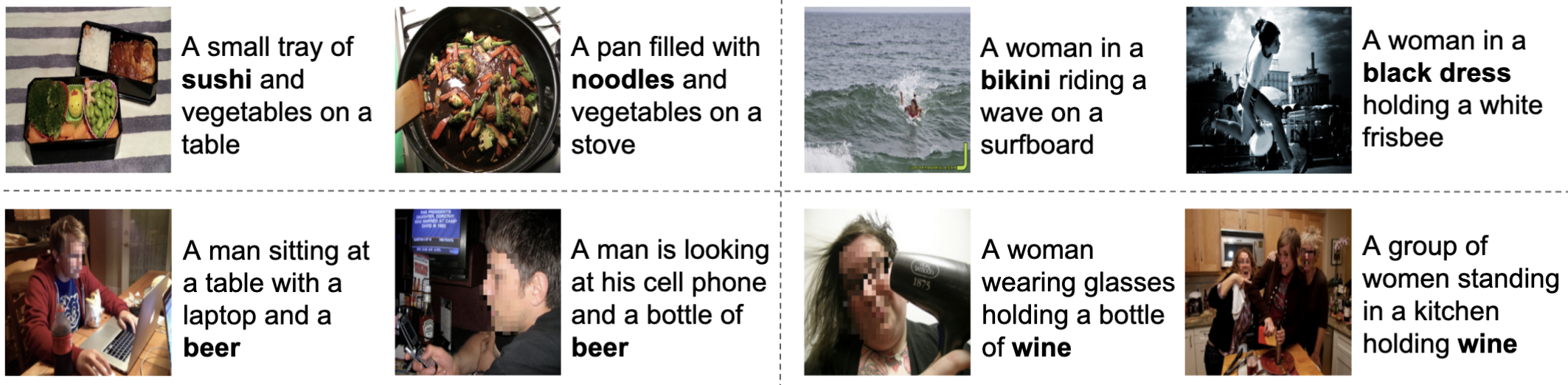}
    \caption{Examples of system-generated captions for COCO that are
    likely explained by stereotypes (i.e., cases that are highly
    ranked according to our heuristic). These cases involve
    hallucinations (i.e., imageable concepts that are not depicted in
    the images) in bold.\looseness=-1}
    \label{fig:stereotype_fp} \Description{Four quadrants of two
    images each, organized by theme. Top left image one shows food in
    bento boxes and a caption of "a small tray of sushi and vegetables
    on a table" despite no sushi in the image, and image two shows a
    pan of stir-fried vegetables with the caption "a pan filled with
    noodles and vegetables on a stove" but there are no noodles. Top
    right image one shows a surfer in the water where you cannot tell
    what they are wearing with the caption "a woman in a bikini riding
    a wave on a surfboard" and image two is in black and white showing
    someone with a ponytail in a tshirt and shorts with a frisbee and
    the caption "a women in a black dress holding a white frisbee."
    Bottom left image one shows a person working on a laptop captioned
    "a man sitting at a table with a laptop and a beer" but a beer
    cannot be seen; image two shows someone on their phone captioned
    "a man is looking at his cell phone and a bottle of beer" but
    again there is no beer to be seen. Bottom right image one shows
    someone with glasses holding a hair blow drier captioned "a woman
    wearing glasses holding a bottle of wine", and image two shows a
    group of people in a kitchen captioned "a group of women standing
    in a kitchen holding wine" when there is no wine.}
\end{figure*}

We found that 11,328 of the 17,360 system-generated captions for
COCO---that is, 65\%---incorrectly included at least one word in a way
that is consistent with one of the three scenarios described in the
previous paragraph. 23\% of these cases involve non-imageable
concepts, 9\% involve concepts that are too specific, and 68\% involve
hallucinations. We provide examples of cases involving hallucinations
that are likely explained by stereotypes (i.e., cases that are highly
ranked according to our heuristic) in
Figure~\ref{fig:stereotype_fp}. Although 11,328 is a large number, we
emphasize that this is best viewed as an overinclusive, upper bound
that is likely influenced by assumptions that do not hold. For
example, it is likely that the human-generated labels are not, in
fact, high quality, meaning that many cases involving concepts that
are too specific or hallucinations are not genuine false positives. In
addition, WordNet's hyponym--hypernym relationships do not always
reflect colloquial uses of language. For instance, \word{couple} is
considered a hyponym of \word{group}, while \word{street} is
considered a hyponym of \word{road}. These words occur in many of the
cases involving concepts that are too specific, although they are not
typically used in ways that reflect these relationships. We similarly
found that 11,539 of the 14,560 system-generated captions for
CC---that is, 79\%---incorrectly included at least one word in a way
that is consistent with the three scenarios described above. Again, we
emphasize that this is best viewed as an overinclusive, upper
bound.\looseness=-1

\subsubsection{Captions that differ in the objects that are correctly mentioned}
\label{sec:stereotyping_true_positives}

We hypothesized that after controlling for the size and location of
the objects depicted in images, any differences between social groups
in the objects that are correctly mentioned in captions (i.e., true
positives) may be explained by stereotypes. To measure these
differences, we drew on the work
of \citet{berg2012importance}. Because this technique requires
demographic annotations, we could only use it to obtain measurements
for COCO. We were also restricted to considering only those social
groups reflected in the demographic annotations collected
by~\citet{zhao2021caption}---that is, male and female (perceived
binary gender) and darker and lighter (skin tone).\looseness=-1

\begin{figure*}[t!]
  \centering \includegraphics[width=\linewidth]{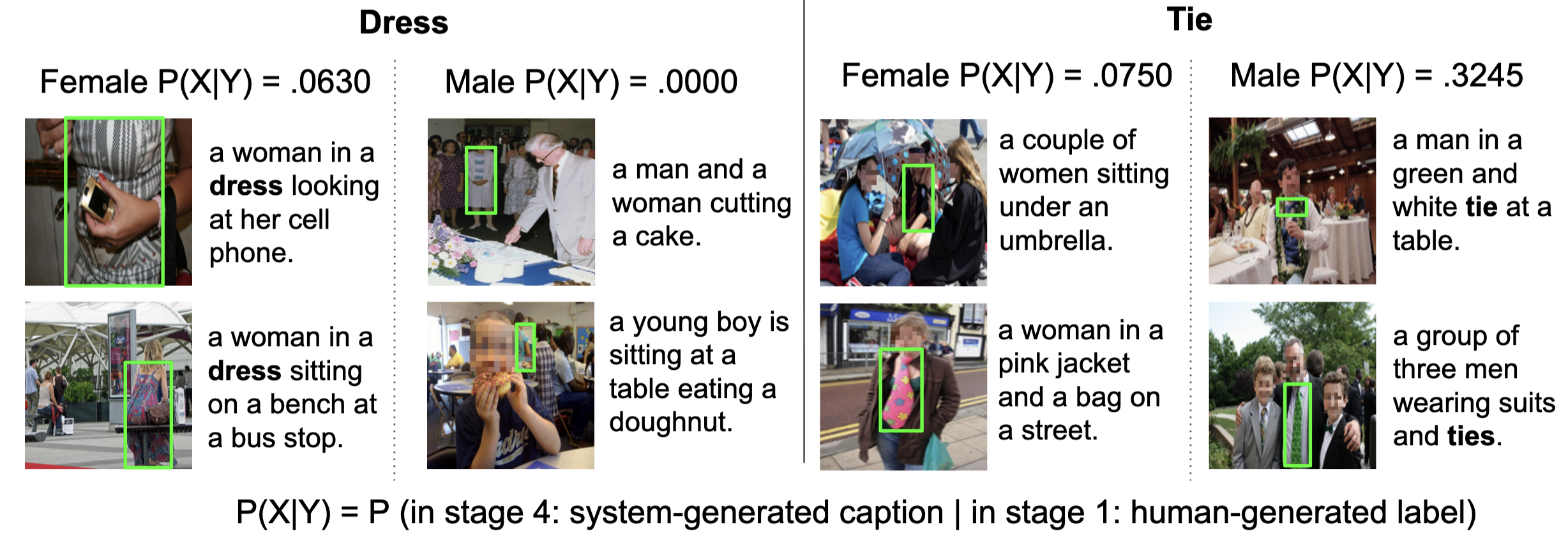} \caption{Examples
  of system-generated captions for COCO where (left) dresses worn by
  people in the background are, perhaps rightfully, not mentioned when
  the people in the foreground are labeled as male and where ties worn
  by people in the foreground are not mentioned when the people in the
  foreground are labeled as female, but are mentioned when those
  people are labeled as
  male.\looseness=-1} \label{fig:stereotype_tp_gen} \Description{Left
  side shows qualitative examples of dress. Probability that dress is
  in the middle caption given that it's in the human label for the
  female label is .0630, and for the male label is .0000. Images for
  female label show the central person in a dress, and images for the
  male label show someone in the background who is not the focus of
  the image wearing a dress. Right side shows the same but for
  tie. The probability for female label is .0750 and for male label is
  .3245. All images show a person who is a focal part of the image
  wearing a tie, regardless of female or male label.}
\end{figure*}

For each pair of social groups (i.e., male and female or darker and
lighter), we treated stage 1 (human-generated labels) as ``ground
truth'' and focused on only the 500 most common object types across
stages 1 and 4 (system-generated captions). For each object type, we
first selected the images that depict an object of that type according
to the human-generated labels and that also depict a person belonging
to either social group according to the demographic annotations
collected by~\citet{zhao2021caption}. We then labeled each image so as
to indicate whether an object of that type is also mentioned in its
system-generated caption---that is, whether the image is a true
positive or a false negative. Next, we created 1,000 train--test
splits of the images, using 70\% for training and 30\% for testing. If
more than 900 of these splits yielded training datasets that contained
both true positives and false negatives, we fit a set of logistic
regression models for that object type---one for each train--test
split where the training dataset contained both true positives and
false negatives. Each model had 1,001 features, where the first 1,000
features were the sizes and locations of the 500 most common object
types, including this one, thereby controlling for the size and
location of all objects of those types. The last feature was the
social group (e.g., male or female) of the largest person depicted in
the image according to the demographic annotations collected
by~\citet{zhao2021caption}. The coefficient for this feature captures
any difference between social groups in true positives for that object
type; we used the set of logistic regression models to obtain
confidence intervals for this coefficient. Having fit a set of
logistic regression models for each of the 500 most common object
types across stages 1 and 4, we restricted our focus to only those
object types whose 95\% confidence interval for this coefficient did
not include zero. This left 23 object types when considering gender
and 20 when considering skin tone. To facilitate interpretation, we
ranked these object types by their models' average accuracies for
their test datasets. We found, for example, statistically
significantly fewer captions that correctly include \word{dress} for
people who are labeled as male according to the demographic
annotations collected by~\citet{zhao2021caption} than for people who
are labeled as female. Similarly, we found statistically significantly
fewer captions that correctly include \word{tie} for people who are
labeled as female according to the demographic annotations collected
by~\citet{zhao2021caption} than for people who are labeled as male.
Although we cannot be sure these differences are explained by
stereotypes, Figure~\ref{fig:stereotype_tp_gen} contains examples of
system-generated captions where dresses worn by people in the
background are, perhaps rightfully, not mentioned when the people in
the foreground are labeled as male and system-generated captions where
ties worn by people in the foreground are not mentioned when those
people are labeled as female, but are mentioned when the people in the
foreground are labeled as male.\looseness=-1

\subsection{Demeaning}
\label{sec:demean}

As described in Section~\ref{sec:types of harms}, demeaning occurs
when social groups are cast as being lower status and less deserving
of respect~\cite{katzman2021tagging}. We propose four techniques for
measuring demeaning. The first technique focuses on cases where words
that are known to be demeaning are included in captions. The second,
third, and fourth techniques capture facets of demeaning that are
unique to image captioning and do not occur in the context of object
detection or image tagging: the second focuses on differences between
social groups in whether people depicted in images are mentioned in
captions, the third focuses on cases that involve particular
context-specific demeaning harms (e.g., calling Black
men \word{boys}), and the fourth focuses on cases where captions use
identity adjectives as nouns (e.g., ``the female walked into the
room'').\looseness=-1

\subsubsection{Captions that include demeaning words}
Although our first measurement technique is conceptually simple, it is
particularly challenging to implement because image captioning systems
typically use open-ended vocabularies. As a result, it is impossible
to manually examine all possible words that could be generated to
determine which ones are indeed demeaning. To address this challenge,
we relied on two word lists---one for objects and one for
attributes. For objects, we used the offensive synsets in the people
subtree of WordNet, as in previous research~\cite{yang2020filter}; for
attributes, we used those adjectives in the judgment category
of \citet{vanmiltenburg2018people}. This technique rests on the
assumption that WordNet's assignments of words to synsets are correct
and the assumption that each word's most common synset is the right
one to use. Because these assumptions may not always hold, we made
three measurements for each word mentioned in a caption:\looseness=-1
\begin{itemize}
    \item Lower bound: if every synset the word belongs to is in one of the demeaning word lists.
    \item Estimate: if the word's most common synset is in one of the demeaning word lists.
    \item Upper bound: if any synset the word belongs to is in one of the demeaning word lists.
\end{itemize}

\begin{figure*}[t!]
  \centering
  \includegraphics[width=.8\linewidth]{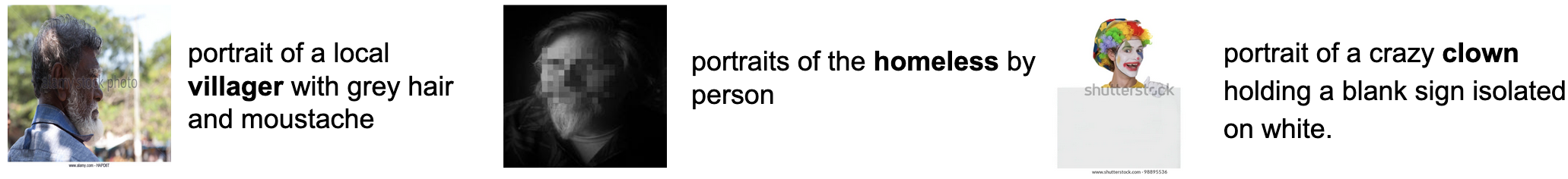}
  \caption{Example of system-generated captions for CC containing words that are known to be demeaning in bold.}
  \label{fig:demean_offensive}
  \Description{Three images, where the captions and corresponding bolded word is "portrait of a local villager with grey hair and moustache" and "villager", then "portraits of the homeless by person" and "homeless", then "portrait of a crazy clown holding a blank sign isolated on white" with "clown."}
\end{figure*}

We found that none (lower bound zero, upper bound 977) of the
system-generated captions for COCO include words that are known to be
demeaning. Meanwhile, we found that 28 (lower bound seven, upper bound
613) of the system-generated captions for CC contain words that are
known to be demeaning. We provide examples of these captions in
Figure~\ref{fig:demean_offensive}. Interestingly, we found that 58
(lower bound 13, upper bound 2,492) of the human-generated captions
for COCO include words that are known to be demeaning, while 37 (lower
bound 11, upper bound 662) of the human-generated captions for COCO
include words that are known to be demeaning. In other words, for both
datasets, VinVL generates captions that include fewer demeaning words
than the human-generated captions. This is likely because
state-of-the-art image captioning systems, including VinVL, generate
less diverse captions than humans, as mentioned in
Section~\ref{sec:training}\looseness=-1

\subsubsection{Captions that differ in whether people depicted in images are mentioned}
We hypothesized that a failure to mention people depicted in images is
demeaning because it is a form of
dehumanization~\cite{haslam2007dehumanization,
bastian2011dehumanization, barlas2021dehumanization}. Our second
measurement technique therefore focuses on differences between social
groups in whether people depicted in images are mentioned in
captions. Because this technique requires demographic annotations, we
could only use it to obtain measurements for COCO and not for CC. We
restricted our focus to those images where people bounding boxes cover
more than 10\% of the image, thereby excluding images in which there
are people depicted in the background who are genuinely not worth
mentioning. For each pair of social groups (i.e., male and female or
darker and lighter), we treated stage 1 (human-generated labels) as
``ground truth'' and assessed whether people belonging to those social
groups according to the demographic annotations collected
by~\citet{zhao2021caption} are mentioned in the system-generated
captions, calculating the fraction of images in which they are not
mentioned. We found, for example, that the 95\% confidence interval
for the difference between darker and lighter was $.0113\pm.0157$. We
repeated these steps for stage 2 (system-generated labels) and stage 3
(human-generated captions), again treating stage 1 as ``ground
truth,'' and found that their 95\% confidence intervals were
$.0022\pm.0067$ and $.0003\pm.0069$, respectively. None of these
differences are statistically significant, although we note that the
human-generated captions yielded the smallest difference between
darker and lighter, while the system-generated captions yielded the
largest, suggesting that the system as a whole may be amplifying
demeaning harms~\cite{wilson2019inequity}.\looseness=-1

\begin{figure*}[t!]
  \centering
  \includegraphics[width=\linewidth]{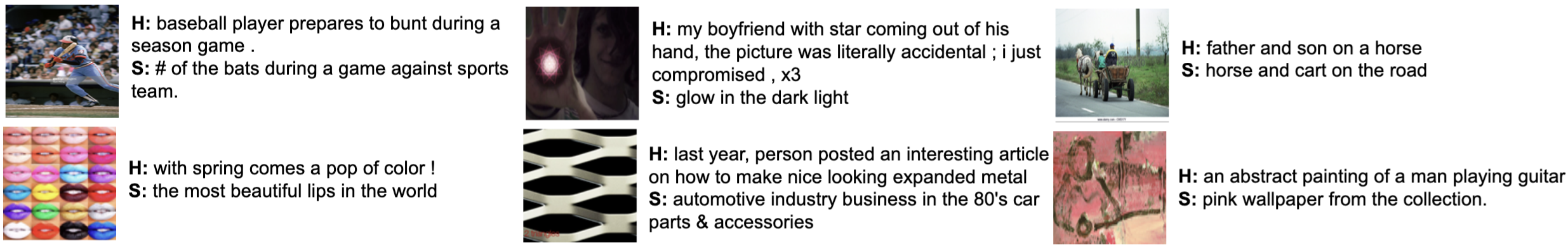}
  \caption{Examples from CC where people mentioned in the human-generated captions (denoted by the prefix \textbf{H}) are not mentioned in the system-generated captions (denoted by the prefix \textbf{S}). The top three images do indeed
depict people who are not mentioned in the system-generated
captions, while the bottom three images reflect different ways
in which our assumptions do not hold.}
  \label{fig:demean_ignore_cc}
  \Description{The top row contains three images with people in them, and the corresponding captions are "H: baseball player prepares to bunt during a season game; S: # of bats during a game against sports team", then "H: my boyfriend with star coming out of his hand, the picture was literally accidental; I just compromised, x3; S: glow in the dark light" then "H: father and son on a horse; S: horse and cart on the road". The bottom row contains three images without any people in them, and the corresponding captions are "H: with spring comes a pop of color!; S: the most beautiful lips in the world", then "H: last year, person posted an interesting article on how to make nice looking expanded metal; M: automative industry business in the 80's car parts & accessories", then "H: an abstract painting of a man playing guitar; S: pink wallpaper from the collection."}
\end{figure*}

Because we do not have access to human-generated labels for CC, we
instead treated stage 3 (human-generated captions) as ground truth and
assessed whether people mentioned in the human-generated captions are
also mentioned in the system-generated captions, regardless of their
social groups, calculating the fraction of images in which they are
not mentioned. Although we found some cases in which people mentioned
in the human-generated captions are not mentioned in the
system-generated captions---that is, possible demeaning harms---we
also found that there are many cases in which our assumptions do not
hold, leading to incorrect measurements. In
Figure~\ref{fig:demean_ignore_cc}, the top three images do indeed
depict people who are not mentioned in the system-generated
captions. However, the bottom three images all reflect different ways
in which our assumptions do not hold. In the first, our assumption
that each word's most common synset is the right one to use does not
hold because \word{pop} has been converted to the synset that
represents a father. As a result, it appears as if the human-generated
caption mentions a person, although this is not the case. In the
second image, our assumption that the human-generated captions are
high quality does not hold because the human-generated caption for
this image refers to the person who posted the image. In the third
image, our assumption that there is a ``correct'' way to caption an
image does not hold because the image is an abstract painting that
allows for many reasonable interpretations.\looseness=-1

\subsubsection{Captions that involve context-specific demeaning harms}
Measuring the extent to which captions involve particular
context-specific demeaning harms is challenging because of the amount
of contextual and historical knowledge required to identify such
harms. As a result, our third measurement technique requires human
input and cannot be fully automated. Drawing on previous research and
recent situations where such harms have been caused by systems
deployed in the real world, we focus on four context-specific
demeaning harms: the first is calling Black
men \word{boys}~\cite{ldf2010case}, the second is calling
women \word{girls}~\cite{lerner1976girls}, the third is incorrectly
mentioning a weapon in the caption for an image of someone who is
Black,\footnote{See \url{https://algorithmwatch.org/en/google-vision-racism/}. We
note that this is also a stereotyping harm, as mentioned in
Section~\ref{sec:stereotype}.} and the fourth is calling Black people
animals~\cite{guynn2015photos}. By focusing on these harms, we do not
intend to overemphasize demeaning harms that are already widely known,
but rather to demonstrate how to leverage existing
knowledge.\looseness=-1

For COCO, we identified cases where captions involve context-specific
demeaning harms by treating stage 1 (human-generated labels) as
``ground truth.'' Because we do not have access to human-generated
labels for CC, we treated stage 3 (human-generated captions) as
``ground truth.'' We found that 27 of the system-generated captions
for COCO and 45 of the system-generated captions for CC involve one or
more of the four context-specific demeaning harms described
above. Interestingly, we found that 178 of the human-generated
captions for COCO involve one or more of these context-specific
demeaning harms. For both datasets, calling
women \word{girls}~\cite{lerner1976girls} is more prevalent than the
other three harms described above, although this may be because there
are substantially fewer images of Black people than there are images
of women.\looseness=-1

\subsubsection{Captions that use identity adjectives as nouns}

\begin{figure*}[t!]
  \centering
  \includegraphics[width=\linewidth]{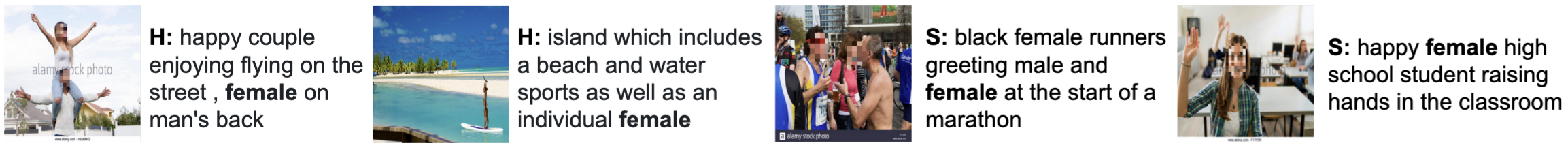}
  \caption{Examples of human-generated captions (\textbf{H}) and system-generated captions (\textbf{S}) for CC that use \word{female} to describe a woman.}
  \label{fig:demean_noun}
  \Description{Four images with people in them with the following captions: "H: happy couple enjoying flying on the street, female on the man's back", "H: island which includes a beach and water sports as well as an individual female", "S: black female runners greeting male and female at the start of a marathon," and "S: happy female high school student raising hands in the classroom."}
\end{figure*}

Using an identity adjective as a noun (e.g., ``the female walked in
the room'') is demeaning because it reduces the person in question to
that aspect of their identity (e.g., their reproductive ability). Our fourth measurement technique
therefore focuses on cases where captions use identity adjectives as
nouns. We restricted our focus to the use
of \word{female} to describe a woman. We used NLTK's part-of-speech
tagger~\cite{bird2009nltk} to identify such cases. We found that none
of the system-generated captions for COCO involve the use
of \word{female} to describe a woman, while two of the
system-generated captions for CC involve the use of \word{female} to
describe a woman. Interestingly, we found that 33 of the
human-generated captions for COCO involve the use of \word{female} to
describe a woman, while nine of the human-generated captions for CC
involve the use of \word{female} to describe a woman. This is likely
because VinVL generates less diverse captions than
humans. Figure~\ref{fig:demean_noun} contains examples of
human-generated captions (\textbf{H}) and system-generated captions
(\textbf{S}) for CC that use \word{female} to describe a woman. The
first example is notable because it uses \word{man}
and \word{female}, rather than \word{man} and \word{woman}. In the
last example, the part-of-speech tagger has incorrectly
tagged \word{female} as a noun.\looseness=-1

\section{Discussion}
\label{sec:discussion}
\subsection{Reflections}

In this paper, our goal was not to audit a particular image captioning
system (in this case VinVL) but rather to develop appropriate
measurement techniques for doing so, in turn providing us with an
opportunity to reflect on the many challenges involved. Despite our
best efforts to develop normatively grounded measurement techniques
that are well-tailored to the unique characteristics of image
captioning, our analysis demonstrates that this is a very difficult
task and that numbers never tell the full story. We did not find
evidence of any particularly surprising representational harms,
although we note that this may be because of the coverage of COCO and
CC. For example, if COCO contains no images that depict a particular
scenario, then the measurements we obtained using COCO reveal nothing
about the captions an image captioning system would generate for
images that do depict that scenario. We did, however, show what it
looks like to attempt to measure representational harms caused by
image captioning systems. This leads us to argue that developing
measurement techniques should be an iterative process that explores
the various ways that different types of harms can manifest. Ideally
this iterative process would be participatory, incorporating the lived
experiences of people who have been or could be harmed by image
captioning systems, and we suggest this as an important avenue to
explore in future work. Finally, we emphasize that the real world is
messy and representational harms are not defined by categorical maxims
but rather by nuanced, extrinsic factors that reflect historical
disparities. As a result, any measurement approach necessarily
involves making assumptions that may not always hold, thereby
threatening the validity and reliability of the resulting
measurements. We therefore aimed to be as transparent as possible
about our assumptions throughout.\looseness=-1

\subsection{Potential Mitigation Techniques}
\label{sec:mitigation}

Our measurement approach depends on a framework of four stages,
depicted in Figure~\ref{fig:scenegraph_tiers}. By treating different
stages as ``ground truth'' for one another, we can better understand
where harms arise, in turn enabling us to understand which mitigation
techniques might be most effective. Below we describe some of the
mitigation techniques suggested by our analysis. Just as many of our
measurement techniques cannot be fully automated, the same is true for
these mitigation techniques.\looseness=-1

First, if we find evidence of a harm at stage 3 (human-generated
captions), treating stage 1 (human-generated labels) as ``ground
truth,'' then one possible mitigation technique would be to obtain new
human-generated captions and retrain the system. However, we note that
for this mitigation technique to be effective, it should be undertaken
with care.


Second, if we find evidence of a harm at stage 4 (system-generated
captions), treating any other stage as ``ground truth,'' then
mitigation techniques that target the natural language model are worth
exploring. In some cases, such as where words that are known to be
demeaning are included in system-generated captions, it may be
possible to remove or replace parts of the captions. However, this
technique is challenging to implement for both technical and normative
reasons. Although adjectives can be removed or replaced without
breaking the grammar of a sentence, removing or replacing other parts
of speech may require the sentence to be rewritten. In addition,
replacing a word is a non-trivial task that can be done during
training, during inference, or as a postprocessing step, each of which
has different pros and cons. We also note that using word lists to
determine which words to remove or replace can cause erasing
harms. This is because some words are only harmful in particular
contexts or when used by particular people (e.g.,
\word{twink})~\cite{bender2021parrots}. Removing or replacing these
words means they cannot be used at all. This challenge is well
discussed in the literature on hate
speech~\cite{sap2019hatespeech,macavaney2019hatespeech,
  dodge2021crawl}.\looseness=-1


Third, if we find evidence of a harm at either stage 2
(system-generated labels) or stage 4 (system-generated captions), then
mitigation techniques involving changes to the computer vision model,
the natural language model, the system as a whole, or the training
approach may be effective. This is particularly appropriate when it is
not possible to obtain new human-generated labels or human-generated
captions due to external constraints~\cite{jo2020archives}. Depending
on the type of harm to be mitigated, possible changes include focusing
on the correct parts of an image~\cite{liu2017attention}, being less
susceptible to spurious correlations~\cite{wang2019balance}, being
better at handling long-tailed label
distributions~\cite{tang2020longtail}, and generating more diverse
captions~\cite{luo2018discriminability}.\looseness=-1

Finally, some mitigation techniques are unique to particular
harms. For example, when mitigating context-specific demeaning harms,
it is possible to raise the threshold for mentioning animals when an
image also depicts people~\cite{wu2017fb}. That said, we caution
against developing mitigation techniques that are narrowly targeted at
one particular technique for measuring a harm unless other measurement
techniques are also used to assess those mitigation techniques'
effectiveness.\looseness=-1


\section{Conclusion}
\label{sec:conclusion}
In contrast to previous research, which has largely considered the
fairness of image captioning systems through the underspecified lens
of ``bias,'' we presented a set of techniques for measuring five types
of representational harms caused by image captioning systems, as well
as the resulting measurements obtained for COCO~\cite{chen2015cococap}
and CC~\cite{sharma2018cc} using the VinVL image captioning
system~\cite{zhang2021vinvl}. Throughout, we discussed the assumptions
underlying our measurement approach and pointed out when they did not
hold. We demonstrated that developing normatively grounded measurement
techniques that are well-tailored to the unique characteristics of
image captioning is a very difficult task. That said, we emphasize
that we must resist the temptation to measure only those properties or
behaviors that are easy to measure.\looseness=-1


\begin{acks}
We thank Sarah Bird, Zhe Gan, Sunnie S. Y. Kim, Anne
Kohlbrenner, Vivien Nguyen, Lijuan Wang, and Zeyu Wang for their
feedback, as well as members of the FATE group at Microsoft
Research. This work was partially supported by National Science Foundation
Graduate Research Fellowship \#2039656 to Angelina Wang. Any opinions,
findings, and conclusions or recommendations expressed in this
material are those of the authors and do not necessarily reflect the
views of the National Science Foundation. This work was also partially
supported by Microsoft. Angelina Wang was an intern at Microsoft
Research while undertaking parts of this work; the other authors are
employees of Microsoft.\looseness=-1
\end{acks}

\bibliographystyle{ACM-Reference-Format}
\bibliography{bib}




\end{document}